\documentstyle[12pt]{article}
\newcommand\be{\begin{eqnarray}}
\newcommand\ee{\end{eqnarray}}

\begin{document}

\begin{tabbing}
\`SUNY-NTG-92-37\\
\`Nov. 1992
\end{tabbing}
\vbox to  0.8in{}
\centerline{\Large \bf Dilepton
and photon production }
\centerline{\Large \bf in the 'hot-glue' scenario }
\vskip 2.5cm
\centerline{\large  E.Shuryak and L.Xiong }
\vskip .3cm
\centerline{Department of Physics}
\centerline{State University of New York at
Stony Brook}
\centerline{Stony Brook, New York 11794}
\vskip 0.35in
\centerline{\bf Abstract}
\indent
 Perturbative analysis of parton kinetics for high
energy nuclear collisions
  shows, that  thermal equilibration
of gluons  happens very fast, while
quark production is much slower.
A simple 'minimal model' is proposed which includes only
in-coming quarks and anti-quarks.
We have found that  a
 smaller quark number
 is more than compensated by the fact, that they are imbedded into the
hotter glue. Predicted yield of dileptons and
 photons in the interesting kinematic region
 are larger than considered before, and are
 quite observable
at RHIC and LHC.

\vskip .25cm
\vfil
\noindent
${^\dagger}$Supported in part by the US Department
 of Energy under Grant No. DE-FG02-88ER40388.
\eject

\newpage
\pagestyle{plain}
\setlength{\baselineskip}{22pt}
\vfill\eject

   The main objective of the future experimental heavy-ion program
at RHIC and LHC is the production and studies  of a new form of matter, the
so called {\it quark-gluon
plasma} \cite{Shuryak_80},
 which is expected to exist
during the first few fm/c of the collisions. Extrapolations from lower energies
are difficult, because for heavy ion collisions
   at RHIC and LHC energies one is entering a new dynamical regime, in which
the so called {\it 'semi-hard'}
processes, involving partons with momenta (and momenta transfer)
 $p\sim$ 1-3 GeV,
can no longer be considered as
  isolated rare events, but are involved in some complicated,
cascade-type processes.

  Early discussions of parton kinetics taking place
 {\it after the primary collisions} can be found in
\cite{Shuryak_80}, and that of 'partonic saturation' effects occurring
{\it before the  collision} can be found  in \cite{GLR}.
The role of few-GeV partons (known also as
'mini-jets') in nuclear collisions was discussed in
\cite{Kajantie_minijets,Eskola_etal,Blaizot_Mueller}. Among
important recent  development are estimates of parton
energy losses in QGP (see \cite{Wang} and references
therein), which for the gluons reach
$\tau_g dE/dx \approx 2-3 $  GeV/fm. Thus, contrary
to some earlier claims,  most of the 'mini-jets'
should be trapped in the fireball, created in central collisions,
 and therefore  contribute to
equilibration processes.

  One open question is
 whether the main amount of produced entropy is due to complicated
partonic substructure of
nuclei themselves, and this entropy is just 'liberated'
 at the collision moment, or it is
produced in a more conventional manner, due to  rescattering and production
processes
{\it after} the primary collisions took place.
 {\it Assuming} that
 perturbative QCD and parton model to be applicable down to 1 GeV scale
 (x$\sim $0.01, at RHIC), one  may come to the former conclusion
 \cite{Blaizot_Mueller}. If
 gluonic structure functions are indeed so
 'dense', that  one can speak about QGP as soon as partons
being separated from each other after the first collision,
at the time
scale $\tau_{separation}\sim 1/p_t \sim .2 fm/c$. Consequences of this
(the most optimistic) scenario for dileptons
 were recently discussed in \cite{Kapusta_etal}, it leads to very high initial
temperatures and very big yields.
However, we do not think such regime
can be dynamically sustained (at least, it is not
so if one is using perturbative cross sections).

 The specific 'hot glue' scenario to be discussed below was recently
 suggested by one of us \cite{Shuryak_twostage}.
   An ambitious program to study 'partonic cascades',
  tracing the system
all the way from the
structure functions to equilibrated QGP,
 was recently started
\cite{Geiger_Muller}.
The results of the simulations   have essentially supported the
'hot glue' scenario. In particular, the time during which most of the
entropy is produced was found to be $\tau_{entropy} \approx 1/2$ fm/c
 for RHIC conditions, momenta distribution approach thermal ones,
 and also gluons clearly outnumber quarks.

  For such a new scenario one has to reconsider
all  previously suggested  QGP signatures. As a benchmark,
we use the so called 'standard' scenario, which implies
{\it complete} equilibration of QGP
 by the time $\tau_0= 1 fm/c$.
In this case the estimated  initial temperatures (see details e.g. in
recent review \cite{Satz}) should be
$ T_i \approx 240\, MeV (RHIC), T_i \approx 290\, MeV (LHC)$, to be compared to
much higher initial temperature $T_i=400-500 MeV$ in the 'hot glue' case.
Some consequences of this picture, like the {\it enhanced charm production}
 were already considered in
\cite{Shuryak_twostage}, and in this paper we concentrate on dilepton and
photon production.

   Let us start with considering the initial conditions.
We adopt a relatively conservative point of view (common for most
 'mini-jet'-related works)
that at least 2-3 GeV parton momentum  is needed in order to
apply perturbative QCD.  If so, the incoming partons are  not
yet 'saturated' at RHIC, but gluons are effectively
rescattered and soon become equilibrated. By their
 ``equilibration time"  $\tau_g\sim 1/3-1/2 fm/c$ we mean the
time at which each secondary gluon  has been in average
{\it rescattered once} \cite{Shuryak_twostage}.
  The problem  is that quarks are not sufficient numerous
inside the parton 'sea', and the relevant production
processes are too slow to be operative at few-fermi time scale
\cite{Shuryak_twostage}. Therefore,
we are left with a {\it relatively small quark admixture}
imbedded into the hot (and quickly equilibrated) glue.

   We suggest a 'minimal model',
an obvious {\it lower bound}, including
only  the in-coming  'sea' quarks (and antiquarks)
 from the colliding nuclei.
Using proton structure functions (a particular parametrization
can be found e.g. in \cite{Gluck_etal}), and adding together u,d,s sea quarks
and anti-quarks we evaluate their number as
\be N_{q+\bar q} =2\int_{x_{min}}dx( u_{sea}(x)+d_{sea}(x)
+ s_{sea}(x) )
 \approx 1.1\pm .1  \ee
where the minimal value is dictated by the applicability of perturbative QCD,
and we take $x_{min}=.02$. (Actually this integral
should also be cutoff at some upper limit,
which should be determined
dynamically, excluding high-x partons which can not be 'dragged' to
the central rapidity window  by rescattering, but high momenta sea quarks are
few anyway.). For example,
 for central
AuAu collisions at RHIC
 the  number of  quarks and anti-quark is then
 $N^{AuAu}_{q+\bar q} > 440$, while for
gluons it is $N^{AuAu}_{g} > 800$.

 Looking at the problem 'from the end'  and normalizing gluons to some
prefixed standard
entropy value (our conservative choice is $dN_\pi/dy=1400$
for AuAu central collisions
at RHIC), one finds  that if the gluonic plasma is
 {\it thermally } equilibrated  in   volume $ (\pi R_A^2 \tau_g)$ for
$\tau_g=.3 fm$, its temperature should be $T\approx 500 MeV$
 (see details in \cite{Shuryak_twostage}). The total number of gluons happen to
be comparable to those 'entering' the collision, as estimated above
on the basis of known structure functions.

 Including the minimal quark admixture evaluated above,
 one reduces the initial temperature to
 $T_0= 460 MeV$. However,
the {\it equilibrated QGP} plasma at such conditions would
contain about $N^{AuAu}_{q+\bar q} \sim 1700 $,
which is much larger than the number of quarks which have  entered
the fireball.
 Since in terms of momentum
distribution quarks  are equilibrated, one can
estimate  rates of any quark-related processes by simply substituting
the proper power of the {\it quark suppression factor} $\xi\sim 1/4$
into the known rates, corresponding to the equilibrium plasma.

   Let us now address the
  dilepton production, which
 is traditionally considered to be one
of the best probes for QGP \cite{Shuryak_80}
(see  recent review in \cite{Ruuskanen}).
  In the
dilepton mass 'window' of interest, M=2-5 GeV,
the main production mechanism is just electromagnetic
$q\bar q$ annihilation, which
  has the following
production rate:
\be {dR_{l^+l^-} \over dM}=\xi^2{2\alpha^2 \over 3 \pi^3} M^2 T K_1(M/T)
\ee
where $K_1(x)$ is a Bessel function, and $\xi$ is
the quark suppression factor
discussed above.

   We have evaluated the dilepton yield
in two ways.
One is based on the simplest scale-invariant
 one-dimensional expansion picture, assuming adiabatic 'cooling'
of the matter.
By integrating
the production rate with the 'temperature profile
function' over the temperature as in
\cite{Shuryak_80}, one gets
\be {dN\over dydM}  = \xi^2 3\pi R_A^2 \tau_0^2 T_0^6 \int_{T_i} {dT \over T^7}
{dR_{l^+l^-} \over d^4x}(T)
\ee
(in principle, the upper integration limit is the
critical temperature $T_c$: but it leads to negligible contribution
for high-mass dileptons to be discussed).
The resulting
invariant mass spectrum are presented in Fig.1a,
 the solid curve for the ``hot-glue'' scenario ($\tau_0=0.3 fm, T_0=460 MeV,
\xi=1/4$)
and the dashed curve for the standard scenario
($\tau_0=1 fm, T_0=240 MeV,
\xi=1$).
The total amount of produced dileptons reflect the number of quarks involved,
while the
slopes of the two curves at high masses
reflect the initial temperatures: so it is not
surprising, that the two curves cross. At large masses
the thermal production yield becomes  smaller than that due to direct
Drell-Yan mechanism
of dilepton production, shown by the dotted line.
 One may therefore conclude, that
in the 'window' M=2-3 GeV the new scenario leads to
the dilepton  yield, which  {\it exceed} that for the
 'standard' scenario and is above or comparable with that due to Drell-Yan
process. Although other possible background should also be considered
(e.g. charm decay) before one can definitely conclude whether it is
experimentally observable, the 'hot glue' prediction looks much more promizing.

  In order to check the sensitivity of these conclusions to the
assumed equilibrium momentum distribution and the one-dimensional expansion,
we have also performed a different evaluation of the dilepton yield
using the cascade model. It has different plasma evolution:
 a cascade-driven
expansion is neither strictly
adiabatic  nor a one-dimensional one.
The results are shown in Fig.1b, and
one can see that this more realistic model leads to similar conclusions
in the invariant mass window of interest. The details of those cascade
simulations
will be  published elsewhere.

  We have also  evaluated the  photon production, due to
the Compton-like process $q(\bar q)g\rightarrow q(\bar q)\gamma$
and quark annihilation $q\bar q\rightarrow g\gamma$
\cite{Shuryak_80}, which dominates at large $p_t$. The thermal
rate is (again, see  general discussion and references in \cite{Ruuskanen})
\be E {dR \over d^3p} = (\xi+\xi^2)
{\alpha \alpha_s T^2 \over 6 \pi^2}
exp(-E/T) [log(6E/g^2 T)+1.62] \ee
where we have introduced the quark suppression factor $\xi$
for these two processes, as explained above.
Integrating over the longitudinal momentum and
over the expansion, assuming it to be one-dimensional, we
have evaluated the photon
yield which is shown in Fig.2.  as for dileptons, predictions of
two scenarios, 'standard' and 'hot glue' ones, are shown by the
solid and dashed curves. Those should also be compared to the projected range
of 'direct photon' production (a range between two dash-dotted lines)
 and the background from
$\pi^0\rightarrow \gamma\gamma$ decay (a range between two dotted lines),
which are taken from ref
 \cite{Ruuskanen}. (Note, the latter background can certainly be
experimentally determined and subtracted, and this should be kept in mind while
comparing the curves.)

  Concluding, the difference between two
scenarios for photon production
is quite crucial from experimental point of view: while
predictions of the 'standard' model are hardly observable,
 those corresponding to
the 'hot glue' scenario   can presumably be seen in the 'window'
at $p_t$=3-5 GeV.

  Considering structure functions
 we have not included the so called {\it shadowing} effect,
 the difference between the nuclear and nucleon structure functions.
It is not yet known for sea quarks at sufficiently small x, and
it can somewhat decrease
the results. However, the same factor  appears in Drell-Yan process and in
the quark number in our 'minimal model', so both thermal production and
 direct ones are affected  by similar
factor. Therefore, since their ratio is the most important experimentally, the
 shadowing effect does not affect most of our conclusions.

 There exist
many other  production mechanisms
  at smaller dilepton masses (or smaller photon $p_t$).
In particular, one should be
able to evaluate thermal production from the hadronic gas.
Recently, the  photon production from it was reconsidered
\cite{XSB}, because it was
 found that the process $\pi\rho\rightarrow A_1 \rightarrow
\gamma \pi$ 'outshine' all others. Obviously, this process
significanly contributes
to small mass dilepton production as well. We will present
detailed discussion of these topics, as well as
our cascade calculations elsewhere.

   Recently, we have obtained a
 paper \cite{Geiger_Kapusta}, in which dilepton yield from the
 non-equilibrium parton cascade is evaluated. Their findings are
similar to ours, with even larger   'window' in which
plasma-related production dominates over other mechnisms.

Summarizing the paper, we have evaluated the dilepton and photon
 spectra at RHIC, based on
 the 'hot glue' scenario.
We have found, that smaller quark fraction does not  make
these signals less useful. On the contrary,
 in the kinematical regions of interest the higher
initial temperature happen to be more important, so
the yields are actually {\it enhanced}.  If so, an experimental
  observation of thermal
dileptons and photons from the quark-gluon
 plasma  have better chances to succeed.

\vskip 1cm
\centerline{Acknowledgements}
\vskip 1cm
This work is partially supported by the
US Department
 of Energy under Grant No. DE-FG02-88ER40388.
\bibliographystyle{try}
\bibliography{ref}

\newpage
\centerline{Figure caption}
1.The invariant mass spectrum of dilepton production from quark-antiquark
annihilation during the plasma expansion
in Au+Au collisions at 100+100 GeV/nucleon, calculated with
(a) the hydrodynamical scale-invariant one-dimensional expansion model;
and (b) the parton cascade simulations.
The  curves are explained in the text.

2.The transverse momentum distribution of photons, produced from quark-gluon
plasma. The curves are explained in the text.
\end{document}